# Embedded Machine Learning for Solar PV Power Regulation in a Remote Microgrid

Yongli Zhu
School of System Science and Engineering
Sun-Yat Sen University
Guangzhou, China
yzhu16@alum.utk.edu

Linna Xu
School of System Science and Engineering
Sun-Yat Sen University
Guangzhou, China
xuln6@mail2.sysu.edu.cn

Jian Huang
School of System Science and Engineering
Sun-Yat Sen University
Guangzhou, China
huangj629@mail2.sysu.edu.cn

*Abstract*—This paper presents a machine-learning study for solar inverter power regulation in a remote microgrid. Machine learning models for active and reactive power control are respectively trained using an ensemble learning method. Then, unlike conventional schemes that make inferences on a central server in the far-end control center, the proposed scheme deploys the trained models on an embedded edge-computing device near the inverter to reduce the communication delay. Experiments on a real embedded device achieve matched results as on the desktop PC, with about 0.1ms time cost for each inference input.

## I. Introduction

In modern power systems, microgrids have played an important role in renewable energy integration. Solar photovoltaic (PV), as a clean and renewable energy source, has been widely used in microgrids. However, the stochastic and intermittent nature of solar PV power generation poses many challenges to the stable operation of microgrids. To effectively cope with these challenges, developing and applying advanced control and prediction techniques are particularly critical.

There are urban microgrids and remote microgrids. The remote microgrid is typically located in villages, mountains, and islands. The remote microgrid is usually built in a rural community with a small daily loading around noon. Thus, it can encounter severe voltage issues: when the solar irradiation is strong (e.g., from 12 AM to 4 PM), the terminal bus voltages of some inverters will frequently violate the voltage upper limit (e.g., reaching as high as 245V RMS value in a 220V AC system). Then, the microgrid or inverter owners must command those inverters to cease working. Then, around sunset, the ceased inverters can be (sometimes manually) restarted. This issue causes a significant waste of solar energy, delays the microgrid project's return on investment (ROI), and hinders potential residents' willingness to purchase PV inverters. Therefore, the power outputs of PV inverters should be regulated to satisfy the microgrid's voltage requirement.

Since the remote microgrid is typically located in an underdeveloped area, conventional centralized power-control architectures have the following limitations: 1) the microgrid's detailed network topology and line parameters can be unknown; hence, power flow-based methods cannot be applied; 2) the local area's communication infrastructure is typically weak; hence the control-action communication between the microgrid and the far-end regional control center can be unreliable; 3) the PV inverter's key parameters are not available, and its internal control-loop is typically disabled for end-users due to safety and confidential concerns; hence it is hard for the end-users to implement droop control or other advanced control strategies.

To overcome the first limitation, data-driven methods such as machine learning (ML) can be considered. For example, In [1], a graph machine learning model is invented to predict the PV power in multiple sites. In [2], the XGBoost (E**X**treme **G**radient **B**oosting) method is employed for solar PV power prediction and performs well. In [3], the artificial neural network (ANN) is used to characterize the nonlinear relationship between the weather and solar PV power. In [4], the support vector machine (SVM) is adopted to improve the forecasting accuracy for solar PV power.

However, the aforementioned study is mainly based on the centralized scheme, i.e., making inferences on a central control center server and then distributing the commands. Therefore, this paper tries to directly deploy the machine learning model on the edge-computing device (e.g., a low-cost ARM board) in the microgrid. This scheme reduces the time delay caused by the transmitting process of the regulation commands, hence lowering the impact of communication failure between the microgrid and the far-end control center.

In the remaining part of this paper, Section II gives a detailed problem description and settings regarding the PV inverter and microgrid used in this study. Section III presents the training process and results of several machine learning models (for solar PV power regulation) on PC. Section IV introduces the basic idea of embedded machine learning and the deployment results on a real ARM board. The last section concludes the paper with the future work envisioned.

## II. Problem Description

### A. Distributed Solar PV Generators

In a microgrid, solar PV is integrated as a distributed generator (DG) at or near the end-user side through a micro-inverter, typically ranging from 1kW to about 1MW. Due to modern power electronics technologies, the active and reactive power of the micro-inverters can be independently regulated for system needs, e.g., maintaining a qualified terminal voltage

This work is supported by "Scientific Computing and System Modeling: Teaching Reform Project" (45000-51200002)

magnitude. A typical P-Q decoupled control strategy for small-size solar PV inverters is depicted in Fig. 1 [5], where $P_{ref}$ and $Q_{ref}$ are reference values (a.k.a. setting points) for the outer-loop controllers, which are the learning targets of the ML models in this paper. As previously discussed, this paper assumes that the inner-loop controllers are inaccessible.

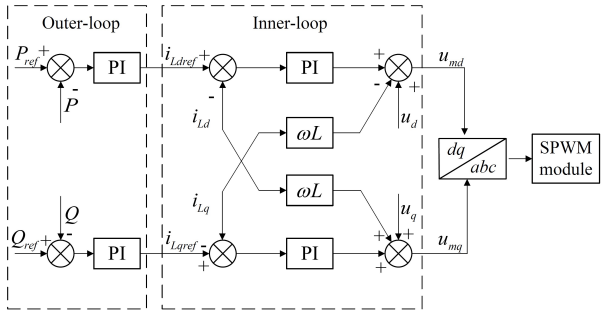

Fig. 1. The diagram of a typical P-Q decoupled control strategy for inverters.

### B. A Real Remote Microgrid

In this study, we obtain real measurement data (voltage, current, power, etc.) of six solar PV inverters from a real microgrid in a remote village in north China, as conceptually illustrated in Fig. 2 (note: the exact network topology and line parameters are not available).

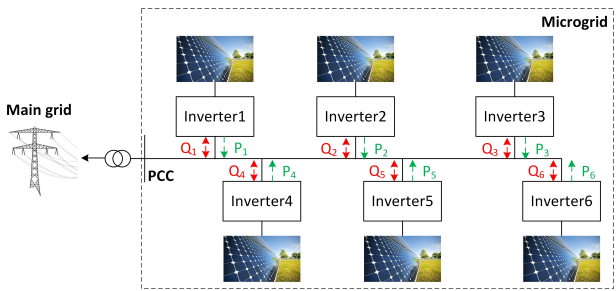

Fig. 2. A conceptual illustration of the studied microgrid in this paper.

Then, we first train and compare several machine models on a desktop PC (Section III) to generate proper setting points respectively for each inverter's active power and reactive power outputs. After that, the trained models are deployed on an embedded ARM board (the edge-computing device) and tested against the results obtained on the PC (Section IV).

### C. Use Machine Learning for PV Power Regulation

In this paper, we use machine learning to establish a mapping between the PV inverter's (active and reactive) power and other quantities (including the terminal voltage). When a desired terminal voltage magnitude is given, the machine-learning model can yield the corresponding power references for the inverter's out-loop controller, as illustrated in Fig. 3.

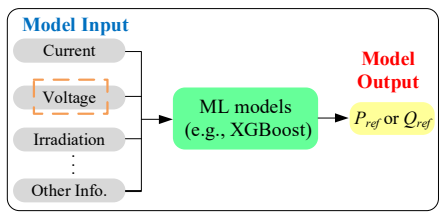

Fig. 3. The basic idea of using machine learning for PV power regulation.

## III. MACHINE LEARNING MODEL TRAINING ON PC

XGBoost is an enhanced decision-tree method that can achieve highly accurate predictions on complex datasets. In this section, the PV active and reactive power models are trained using XGBoost based on the real measurements of the six PV inverters and then compared with two other methods, i.e., SVM and linear regression (LR). All the experiments in this section are implemented on a desktop PC with Intel 5.4GHz CPU and 32GB RAM.

### A. Dataset Description

The dataset used in this paper was collected from a small village in north China, consisting of six PV inverters' data for nearly one month (May to June 2024). The dataset contains three-phase currents and voltages, active and reactive power, and weather information at a 15-minute resolution.

A series of preprocessing works, such as missing-value-filling and dirty-data-cleaning, were performed on the raw dataset to ensure the data quality. Finally, the length of data entries of each PV inverter is between 2000 and 2800.

### B. Machine Learning Models and Performance Metrics

In this paper, XGBoost is chosen as the main method, which is one kind of *ensemble learning* approach. By combining several weak learners (e.g., simple decision trees), it can gradually optimize the performance of the model. It adopts the idea of "boosting": by iterative training, the later model tries to fit the *residual* that the former model failed to handle well to gradually reduce the overall error. Fig. 4 illustrates the basic idea of the XGBoost method.

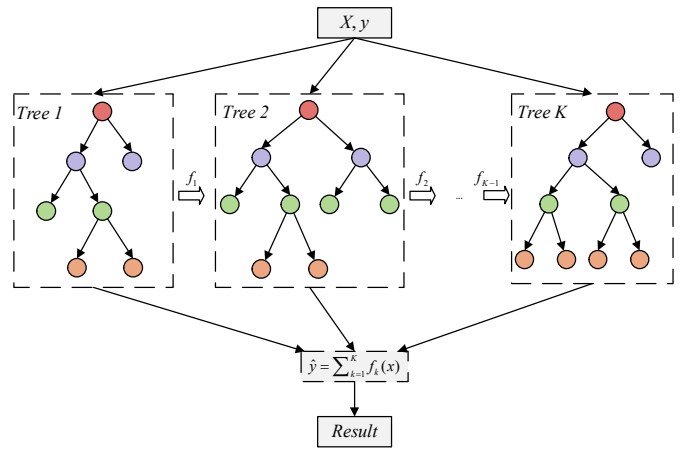

Fig. 4. The basic idea of XGBoost.

The basic idea of XGBoost is to construct an additive model. Assuming that $K$ trees are trained and $n$ is the total number of data samples, then the prediction for the $i$-th sample can be expressed by Eq. (1):

$$\hat{y}_i = \sum_{k=1}^{K} f_k(x_i) \qquad (1)$$

where $f_k$ is a decision tree. Each tree is trained by progressively optimizing the objective function. The gradient boosting algorithm used by XGBoost minimizes the residuals at each iteration. In the $m$-th iteration, the model tries to minimize the objective function in Eq. (2).

$$Obj^{(m)} = \sum_{i=1}^{n} l(y_i, \hat{y}_i^{(m-1)} + f_m(x_i)) + \Omega(f_m) \quad (2)$$

The objective function consists of two parts: the loss function $l$ and the regularization term $\Omega$. The regularization term is shown in Eq. (3).

$$\Omega(f) = \gamma T + \frac{1}{2}\lambda \|\omega\|^2 \quad (3)$$

where $T$ is the number of leaf nodes of the tree, $\lambda$ is the regularization parameter of the weights, and $\gamma$ is the parameter that controls the complexity of the tree. To optimize Eq. (2), XGBoost approximates the objective function using a second-order Taylor expansion as shown in Eq. (4).

$$Obj^{(m)} \approx \sum_{i=1}^{n} [g_i f_m(x_i) + \frac{1}{2} h_i f_m(x_i)^2] + \Omega(f_m) \quad (4)$$

where $g_i$ and $h_i$ are the first-order and second-order derivatives of the loss function, respectively, as defined in Eq. (5) and (6).

$$g_i = \frac{\partial l(y_i, \hat{y}_i^{(m-1)})}{\partial \hat{y}_i^{(m-1)}} \quad (5)$$

$$h_i = \frac{\partial^2 l(y_i, \hat{y}_i^{(m-1)})}{\partial (\hat{y}_i^{(m-1)})^2} \quad (6)$$

By optimizing the above objective functions and continuously adding new trees, XGBoost can build a powerful predictive model that performs well on large-scale datasets with high-dimensional features.

In this study, the feature input of the model is all the data in the dataset other than the active/reactive power, and the prediction label is the active/reactive power. $R^2$ and MAPE are chosen as the performance metric. $R^2$ is a statistical index to measure the total variation of the explanatory variables of a regression model, which indicates the explanatory power of the independent variables on the dependent variable. Its value is in [0,1]. The closer $R^2$ is to 1, the stronger the model's explanatory power is. Its formula is shown in Eq. (7).

$$R^2 = 1 - \frac{\sum_{i=1}^{n}(y_i - \widehat{y_i})^2}{\sum_{i=1}^{n}(y_i - \overline{y})^2} \quad (7)$$

MAPE is another error metric, indicating the percentage of the prediction error to the actual value. Since PV inverters' active and reactive power are dominated by solar irradiation, the dataset contains zero values, which result in infinity when using the conventional MAPE formula. Thus, we use another version of MAPE that can avoid infinity, as shown in Eq. (8).

$$MAPE = (1 - \sqrt{\frac{1}{n}\sum_{i=1}^{n}(\frac{y_i - \widehat{y_i}}{Cap})^2}) \times 100\% \quad (8)$$

where $Cap$ denotes the power capacity of the inverter.

*C. ML Models for Solar PV Reactive Power Regulation*

The training and testing set are randomly divided by the ratio of 8:2. Table I and Table II compare the performance metrics of the three machine learning models for reactive power prediction on the testing dataset.

As can be seen from Table I and Table II, when predicting the reactive power, the $R^2$ of the XGBoost model is closer to 1 compared to the SVM and LR models. The MAPE of the XGBoost model is less than 2% and smaller than the SVM and LR models, which indicates that the XGBoost model's performance is the best.

TABLE I. $R^2$ OF REACTIVE POWER MODELS ON THE TESTING DATA

| Inverter No. | XGBoost | SVM | LR |
|---|---|---|---|
| 1 | **0.9841** | 0.5194 | 0.8427 |
| 2 | **0.9979** | 0.7592 | 0.9354 |
| 3 | **0.9961** | 0.3357 | 0.7860 |
| 4 | **0.9999** | 0.5488 | 0.9989 |
| 5 | **0.9985** | 0.0961 | 0.1777 |
| 6 | **0.9999** | 0.8233 | 0.9892 |

TABLE II. MAPE OF REACTIVE POWER MODELS ON THE TESTING DATA

| Inverter No. | XGBoost | SVM | LR |
|---|---|---|---|
| 1 | **1.69%** | 9.27% | 5.31% |
| 2 | **0.56%** | 6.04% | 3.13% |
| 3 | **0.52%** | 6.80% | 3.86% |
| 4 | **0.05%** | 9.93% | 0.48% |
| 5 | **1.56%** | 38.32% | 36.55% |
| 6 | **0.06%** | 5.00% | 1.24% |

In addition, we apply the above models on a complete day for PV inverter-1, and the results are depicted in Fig. 5. It is clear that the regression curve obtained by XGBoost fits the true value better than the other two methods except for a spike at only one data sample in the rear of the curve. This kind of rear spike prediction can be easily weeded out by a simple threshold logic since there should be no big power output during sunset and evening.

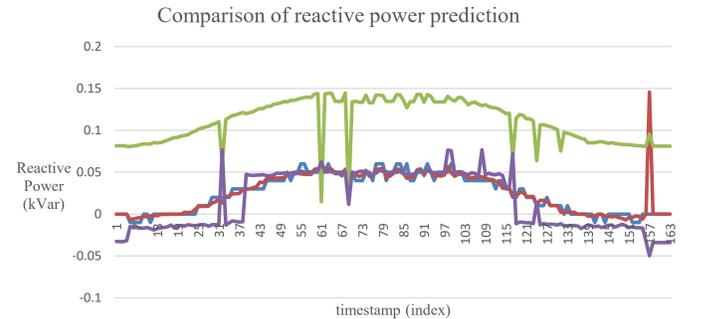

Fig. 5. Comparison of reactive power models' predictions

*D. ML Models for Solar PV Active Power Regulation*

In this section, we train and compare a series of ML models to predict the active power of six PV inverters. Similar to the previous section, comparison results of $R^2$ and MAPE are listed in Tables III and IV. From Table III and Table IV, we can similarly conclude that the XGBoost model performs best in the active power prediction task for the six PV inverters. Thus, the trained XGBoost models will be adopted for deployment on the embedded device in the next section.

TABLE III.    R² OF ACTIVE POWER MODELS ON THE TESTING DATA

| Inverter No. | XGBoost | SVM | LR |
| --- | --- | --- | --- |
| 1 | **0.9907** | 0.6091 | 0.6656 |
| 2 | **0.9929** | 0.6677 | 0.5861 |
| 3 | **0.9963** | 0.5949 | 0.4644 |
| 4 | **0.9869** | 0.4959 | 0.7125 |
| 5 | **0.9982** | 0.8393 | 0.8344 |
| 6 | **0.9995** | 0.7148 | 0.9979 |

TABLE IV.    MAPE OF ACTIVE POWER MODELS ON THE TESTING DATA

| Inverter No. | XGBoost | SVM | LR |
| --- | --- | --- | --- |
| 1 | **2.39%** | 15.54% | 14.38% |
| 2 | **3.44%** | 23.61% | 26.35% |
| 3 | **1.66%** | 17.39% | 20.00% |
| 4 | **3.17%** | 19.72% | 14.89% |
| 5 | **0.94%** | 9.12% | 9.26% |
| 6 | **0.42%** | 10.80% | 0.91% |

A comparison of the active power prediction for a complete day on PV inverter-1 is shown in Fig. 6. We can also observe that the predicted active power curve obtained from the XGBoost model fits the real curve best (nearly overlapped).

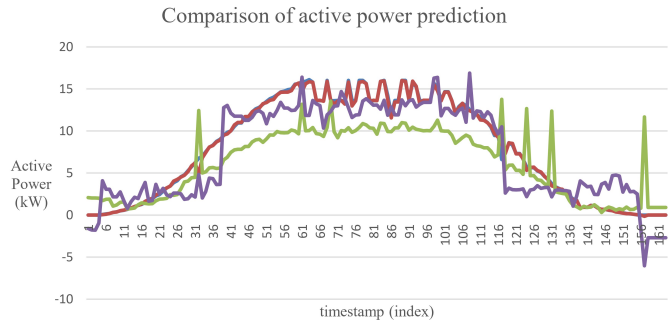

Fig. 6.   Comparison of active power models' predictions

## IV. MACHINE LEARNING MODEL DEPLOYMENT ON ARM

In this section, we will dive into the implementation of deploying the best-performed ML models (i.e., the trained XGBoost models in Section III) on an embedded ARM board.

### A. ARM Board Description

Fig. 7 displays the board where we try to deploy the model. The ARM board in this study is based on ARMv8 architecture, possessing four Cortex-A processors and about 1GB of disk space. It runs an ARM Linux operating system (ver. 4.9.38).

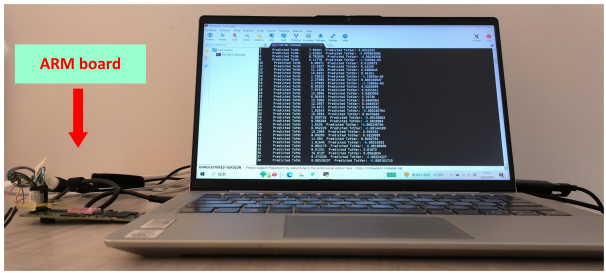

Fig. 7.   The ARM board used in this study.

### B. ONNX Introduction

ONNX (Open Neural Network Exchange) [6] is a neural network exchange format developed by Microsoft for model conversion and exchange. Users can convert models between different machines and operating systems. After exporting trained models to ONNX format, one can use the *onnx-runtime* to execute the model on a different target machine or operating system. The ONNX has been successfully applied in computer vision and edge-computing areas [7]-[9]. As Fig. 8 shows, ONNX consists of an extensive computational graph model supporting a series of built-in operational units and standard data types. Each computational flow graph is defined as a list combined by nodes, and a directed acyclic graph is constructed. Each node owns one or several inputs/outputs and is called an *operational unit*, equivalent to the universal calculation graph that other hardware architectures and operating systems can interpret.

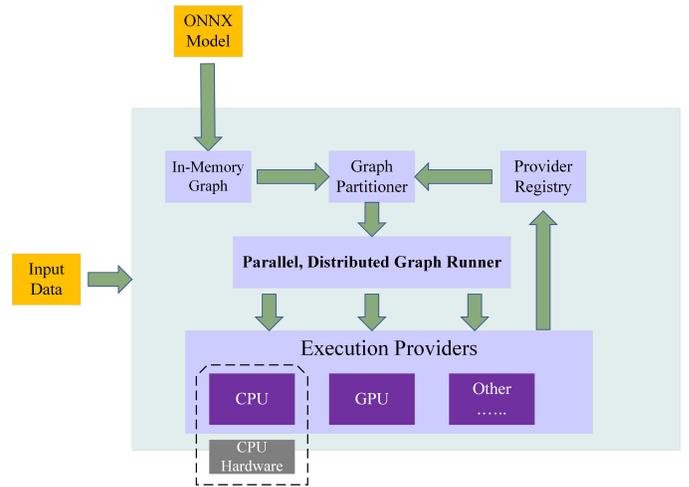

Fig. 8.   The basic architecture and workflow of ONNX.

### C. Deploy Solar PV Reactive Power Models on ARM

In this part, we will present the deployment process of the PV reactive power regulation models and compare the model prediction on the ARM board with that on the PC.

To deploy the model on the ARM board, the machine learning model must first be serialized to the ONNX format that other platforms or programming languages can *invoke*.

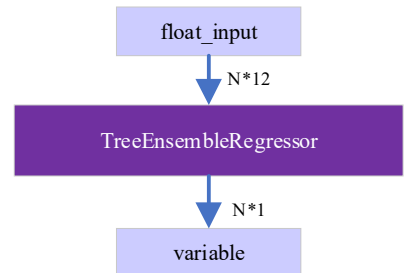

Fig. 9.   Structure of the ONNX-format model.

As shown in Fig. 9, the obtained ONNX-format model has an input that requires a 1x12 vector as input, a *TreeEnsembleRegressor* layer, and an output layer.

```cpp
// invoke the model
float predict_with_onnx(Ort::Session& session, const std::vector<float>& input_data) {
    if (input_data.size() != 12) {
        throw std::runtime_error("Input data must contain exactly 12 elements.");
    }
    std::vector<int64_t> input_tensor_shape = {1, 12}; // input data shape is 1x12
    size_t input_tensor_size = input_data.size(); // total size of input data
    // create input vector
    Ort::MemoryInfo memory_info = Ort::MemoryInfo::CreateCpu(OrtArenaAllocator, OrtMemTypeDefault);
    Ort::Value input_tensor = Ort::Value::CreateTensor<float>(
        memory_info, const_cast<float*>(input_data.data()), input_tensor_size,
        input_tensor_shape.data(), input_tensor_shape.size());
    // set input and output node name
    const char* input_names[] = {"float_input"};
    const char* output_names[] = {"variable"};
    // run the model
    auto output_tensors = session.Run(Ort::RunOptions{nullptr}, input_names,
        &input_tensor, 1, output_names, 1);
```

Fig. 10. Code snippets for invoking the ONNX-format model in C++.

Fig. 10 shows how we invoke the ONNX-format model in C++. To invoke it, we need to create an input vector and an output vector and then use the *onnx-runtime* to call the model. By using a specific cross-compile toolchain to obtain the final executable files, we can then deploy and run those model files on the ARM board.

As an example, Table V shows a portion of predictions yielded by inverter-1's reactive power model (running respectively on PC and ARM). By analyzing 515 test data, We can conclude that the outputs are almost the same, with only a few decimal places apart.

TABLE V.     OUTPUTS OF INVERTER-1'S REACTIVE POWER MODEL ON THE TESTING DATA

| PC | ARM board |
| --- | --- |
| 9.4432897567749 | **9.44329** |
| 9.4616117477417 | **9.46161** |
| 4.76725912094116 | **4.76726** |
| 3.81672215461731 | **3.81672** |
| 7.01548290252685 | **7.01548** |
| 4.76597213745117 | **4.76597** |

Then, we check the identity of the PC's and ARM's outputs by computing the MAPE and RMSE (between the two outputs). From Table VI, they can be considered identical up to about six decimal points. The results here indicate that the machine-learning models have been successfully deployed to our ARM board and can output correct values.

TABLE VI.    PREDICTION RESULTS COMPARISON: PC VS. ARM

| Inverter No. | MAPE | RMSE |
| --- | --- | --- |
| 1 | 0.6896% | 0.000000544789832 |
| 2 | 0.0059% | 0.000000440760592 |
| 3 | 0.0371% | 0.000000354978247 |
| 4 | 0.0121% | 0.000000717906506 |
| 5 | 0.0572% | 0.000003153948804 |
| 6 | 0.0037% | 0.000000948042872 |

### D. Deploy Solar PV Active Power Models on ARM

Here, the results and comparisons of the active power regulation models on PC and ARM are present. Because the model structure of the active power model is similar to that of the reactive power model, hence their converted ONNX-format models have similar structures. Thus, the deployment processes are also similar.

TABLE VII.   OUTPUTS OF INVERTER-1'S ACTIVE POWER MODEL ON THE TESTING DATA

| PC | ARM board |
| --- | --- |
| 3.75673151016235 | **3.75673** |
| 1.91252446174622 | **1.91252** |
| 0.23450544476509 | **0.234507** |
| 9.1589355468750 | **9.15894** |
| 1.61964440345764 | **1.61964** |
| 2.11777901649475 | **2.11778** |

TABLE VIII.  PREDICTION RESULTS COMPARISON: PC VS. ARM

| Inverter No. | MAPE | RMSE |
| --- | --- | --- |
| 1 | 0.0089% | 0.000013004815713 |
| 2 | 0.0554% | 0.000011748417447 |
| 3 | 0.0013% | 0.000016227362974 |
| 4 | 0.0048% | 0.000015246389763 |
| 5 | 0.0041% | 0.000007144940236 |
| 6 | 0.0059% | 0.000012877150695 |

Table VII lists a portion of predictions made on PC and ARM, and Table VIII inspects their differences using MAPE and RMSE. We repeat this experiment on the six inverters. The result indicates that our deployment of the model for active power prediction is also successful.

### E. Time Comparison for Model Inference: PC vs. ARM

Here, the average inference time costs for single data input (executed respectively on PC and ARM) are reported in Table IX. Unsurprisingly, the trained model running on ARM is slower than that on PC since our PC's CPU is much more powerful than the low-cost ARM core. However, the time cost is around 0.1ms, which is fast enough for PV power regulation.

TABLE IX.    COMPARISON OF MODEL INFERENCE TIME: PC VS. ARM

| Inverter No. | Reactive power model | | Active power model | |
| --- | --- | --- | --- | --- |
| | PC | ARM board | PC | ARM board |
| 1 | 0.0072ms | **0.1168ms** | 0.0313ms | **0.1233ms** |
| 2 | 0.0082ms | **0.1122ms** | 0.0373ms | **0.1102ms** |
| 3 | 0.0079ms | **0.1192ms** | 0.0339ms | **0.1249ms** |
| 4 | 0.0094ms | **0.1141ms** | 0.0367ms | **0.1143ms** |
| 5 | 0.0084ms | **0.1208ms** | 0.0323ms | **0.1155ms** |
| 6 | 0.0120ms | **0.1155ms** | 0.0705ms | **0.1165ms** |

## V. Conclusion

In this paper, an embedded machine learning study is conducted to ensure the solar PV inverter performs its power-regulation action quickly and reliably in a remote microgrid. Experimental results on an embedded ARM board show that the deployed model can accomplish the inference task in milliseconds with perfectly matched results against the PC. Future work includes more delicate feature engineering and trying deep learning models.

## References

[1] J. Simeunović, B. Schubnel, P.-J. Alet, R. E. Carrillo, and P. Frossard, "Interpretable temporal-spatial graph attention network for multi-site PV power forecasting," *Appl. Energy*, vol. 327, p. 120127, 2022, doi: https://doi.org/10.1016/j.apenergy.2022.120127.

[2] Phan, Quoc-Thang, Yuan-Kang Wu, and Quoc-Dung Phan. "Short-term solar power forecasting using xgboost with numerical weather prediction." *2021 IEEE International Future Energy Electronics Conference (IFEEC)*. IEEE, 2021.

[3] Chen, Changsong, et al. "Online 24-h solar power forecasting based on weather type classification using artificial neural network." Solar energy 85.11 (2011): 2856-2870.

[4] Ekici, Betul Bektas. "A least squares support vector machine model for prediction of the next day solar insolation for effective use of PV systems." *Measurement* 50 (2014): 255-262.

[5] R. Teodorescu, M. Liserre, P. Rodriguez, Grid Converters for Photovoltaic and Wind Power Systems. West Sussex, UK: Wiley-IEEE Press, 2011.

[6] Open Neural Network Exchange. https://onnx.ai/

[7] J. Moosmann, H. Müller, N. Zimmerman, G. Rutishauser, L. Benini and M. Magno, "Flexible and Fully Quantized Lightweight TinyissimoYOLO for Ultra-Low-Power Edge Systems," in IEEE Access, vol. 12, pp. 75093-75107, 2024.

[8] P. Li, X. Wang, K. Huang, Y. Huang, S. Li, and M. Iqbal, "Multi-Model Running Latency Optimization in an Edge Computing Paradigm," *Sensors*, vol. 22, no. 16, 2022.

[9] K. Desappan *et al.*, "Open source deep learning inference libraries for autonomous driving systems," *Electron. Imaging*, vol. 34, no. 16, pp. 111–118, 2022, doi: 10.2352/EI.2022.34.16.AVM-118.